\documentclass[journal=jacsat,manuscript=article]{achemso}

\usepackage[version=3]{mhchem} 
\usepackage{nicefrac}

\author{Oliver Braun}
\affiliation{Empa, Swiss Federal Laboratories for Materials Science and Technology, Transport at Nanoscale Interfaces Laboratory, Überlandstrasse 129, CH-8600 Dübendorf, Switzerland}
\alsoaffiliation[Basel University]
{Department of Physics, University of Basel, Klingelbergstrasse 82, CH-4056 Basel, Switzerland}

\author{Roman Furrer}
\affiliation{Empa, Swiss Federal Laboratories for Materials Science and Technology, Transport at Nanoscale Interfaces Laboratory, Überlandstrasse 129, CH-8600 Dübendorf, Switzerland}

\author{Pascal Butti}
\affiliation{Empa, Swiss Federal Laboratories for Materials Science and Technology, Transport at Nanoscale Interfaces Laboratory, Überlandstrasse 129, CH-8600 Dübendorf, Switzerland}

\author{Kishan Thodkar}
\affiliation{Empa, Swiss Federal Laboratories for Materials Science and Technology, Transport at Nanoscale Interfaces Laboratory, Überlandstrasse 129, CH-8600 Dübendorf, Switzerland}
\altaffiliation{Current address: Department of Mechanical and Process Engineering, ETH Zurich, Tannenstrasse 3, CH-8092 Zurich, Switzerland}

\author{Ivan Shorubalko}
\affiliation{Empa, Swiss Federal Laboratories for Materials Science and Technology, Transport at Nanoscale Interfaces Laboratory, Überlandstrasse 129, CH-8600 Dübendorf, Switzerland}

\author{Ilaria Zardo}
\affiliation[Basel University]
{Department of Physics, University of Basel, Klingelbergstrasse 82, CH-4056 Basel, Switzerland}

\author{Michel Calame}
\affiliation{Empa, Swiss Federal Laboratories for Materials Science and Technology, Transport at Nanoscale Interfaces Laboratory, Überlandstrasse 129, CH-8600 Dübendorf, Switzerland}
\alsoaffiliation[Basel University]
{Department of Physics, University of Basel, Klingelbergstrasse 82, CH-4056 Basel, Switzerland}
\alsoaffiliation[SNI]
{Swiss Nanoscience Institute, University of Basel, Klingelbergstrasse 82, CH-4056 Basel, Switzerland}

\author{Mickael L. Perrin}
\email{mickael.perrin@empa.ch}
\affiliation{Empa, Swiss Federal Laboratories for Materials Science and Technology, Transport at Nanoscale Interfaces Laboratory, Überlandstrasse 129, CH-8600 Dübendorf, Switzerland}

\title[An \textsf{achemso} demo]
  {Spatially mapping the thermal conductivity of graphene by an opto-thermal method}

\abbreviations{IR,NMR,UV}

\begin{document}


\begin{tocentry}

\includegraphics[width=\linewidth]{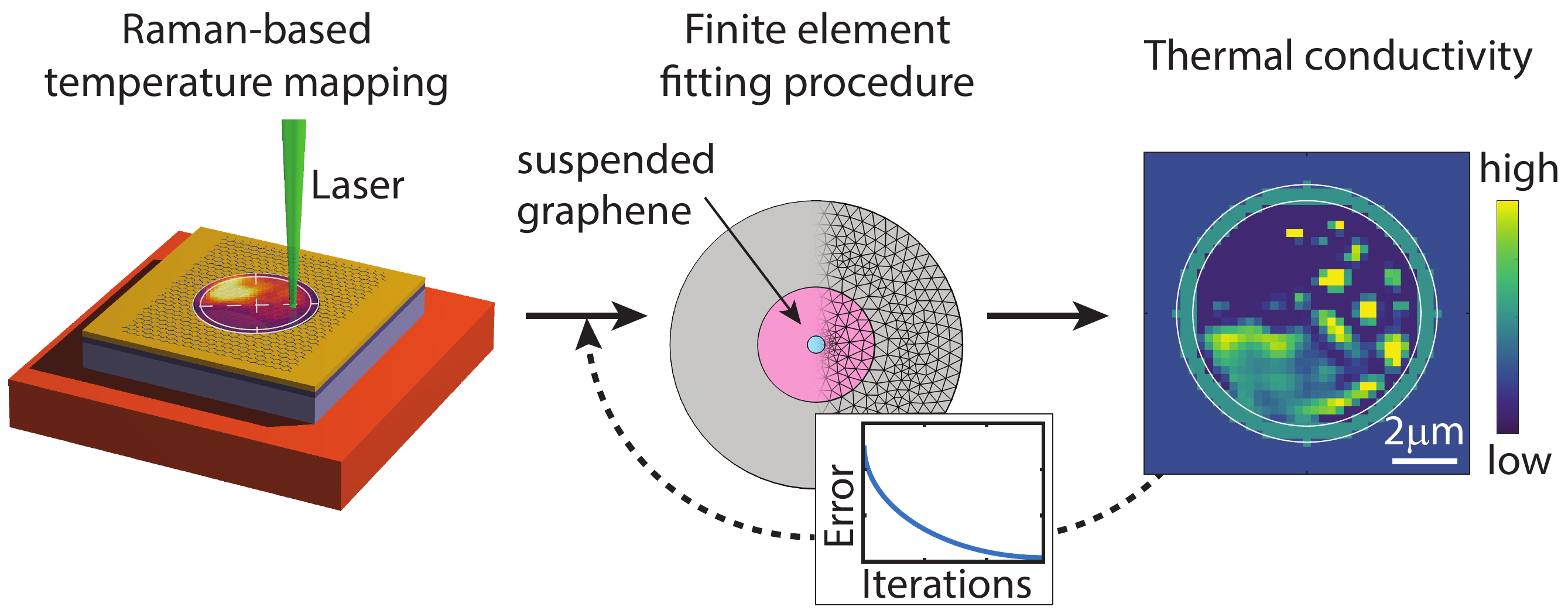}

\end{tocentry}


\newpage
\begin{abstract}
Mapping the thermal transport properties of materials at the nanoscale is of critical importance for optimizing heat conduction in nanoscale devices. Several methods to determine the thermal conductivity of materials have been developed, most of them yielding an average value across the sample, thereby disregarding the role of local variations. Here, we present a method for the spatially-resolved assessment of the thermal conductivity of suspended graphene by using a combination of confocal Raman thermometry and a finite-element calculations-based fitting procedure. We demonstrate the working principle of our method by extracting the two-dimensional thermal conductivity map of one pristine suspended single-layer graphene sheet and one irradiated using helium ions. Our method paves the way for spatially resolving the thermal conductivity of other types of layered materials. This is particularly relevant for the design and engineering of nanoscale thermal circuits (e.g. thermal diodes).

\end{abstract}

\section{Introduction}

Thermal properties of materials are of crucial importance for optimizing heat management in nanoscale devices, with the thermal conductivity as key material property.\cite{Shi.2015,Song.2018} The thermal conductivity is typically determined by monitoring the sample temperature and/or heat flow in response to a local heat source, in combination with an analytical expression or a numerical model. 
For instance, for bulk materials, the well-known 3$\omega$ technique\cite{Corbino.1910,Corbino.1911} is used, while for nanoscale materials, methods such as the thermal bridge method \cite{Seol.2010,Swinkels.2015,Yazji.2016} and micro-Raman spectroscopy\cite{Deshpande.2009,Soini.2010,Reparaz.2014,Neogi.2015} provide the thermal conductivity of the material. 

Of particular interest are the thermal properties of layered materials. Due to their broad range of conductivity values and their atomically thin nature, such materials are highly relevant for heat management at the nanoscale.\cite{Shahil.2012,Balandin.2020} One of the most appealing materials is graphene, with extraordinarily high thermal conductivity values. However, extracting the thermal properties of 2D materials is challenging, in particular when suspended to reduce the influence of the substrate. For example, time-domain thermoreflectance cannot be applied to 2D materials as the material is too thin\cite{Kasirga.2020}. Scanning thermal probe microscopy, on the other hand, despite possessing nanometer resolution, is highly delicate to perform on suspended 2D materials. Moreover, both techniques rely on on-chip heaters for channeling heat into the system. 
Raman spectroscopy can overcome these difficulties, as it can utilize the excitation laser to locally heat the device, while at the same time measuring the local temperature. Moreover, Raman spectroscopy can be conveniently performed on suspended graphene films for eliminating the influence of the substrate. Using Raman spectroscopy, Balandin and Ghosh \textit{et al.}\cite{Balandin.2008, Ghosh.2008} determined the thermal conductivity of suspended graphene to be as high as $\sim$5000~Wm$^{-1}$K$^{-1}$ at room temperature. Their opto-thermal method, measuring the shift of the Raman G-band upon laser irradiation for estimating the local temperature, has been extensively used in literature since. Alternatives based on the intensity ratio of Stokes to anti-Stokes Raman scattering\cite{Faugeras.2010} or the Raman 2D-band\cite{Lee.2011} have also been reported. 

Using this opto-thermal method, the influence of the quality and structure of the graphene, as well as the environment have been extensively investigated. For instance, Cai \textit{et al.}\cite{Cai.2010} reported values for $\kappa$ exceeding $\sim$2500~Wm$^{-1}$K$^{-1}$ for suspended graphene grown by chemical vapor deposition (CVD), and Chen \textit{et al.}\cite{Chen.2011} studied the influence of environment on thermal conductivity of graphene. Isotopically pure $^{12}$C (0.01~$\%$ $^{13}$C) graphene has been shown to exhibit $\kappa$ = 4000~Wm$^{-1}$K$^{-1}$, a factor of two higher than $\kappa$ in graphene composed of a 1:1 mixture of $^{12}$C and $^{13}$C.\cite{Chen.2012} Also, the influence of CVD-graphene's polycrystallinity on the thermal conductivity was studied by Lee \textit{et al.} and Ma \textit{et al.}, revealing that smaller grain sizes drastically reduce $\kappa$ due to grain boundary scattering.\cite{Lee.2017, Ma.2017} Along similar lines, wrinkles\cite{Chen.2012c}, oxygen-plasma induced defects\cite{Zhao.2015}, and electron beam irradiation\cite{Malekpour.2016} have shown to reduce the thermal conductivity.\\

In all the above-mentioned studies, the one-dimensional heat equation is used to fit the experimental temperature and extract the thermal conductivity. However, such approaches yield an average value of thermal conductivity value, not a spatially resolved map. This implied that local variations caused by defects, folds, contaminants, etc. are neglected.\\

For going beyond the average material property value, approaches mapping the temperature distribution in the sample are necessary, combined with multi-dimensional analytical of numerical model. A range of techniques has been developed for nanoscale thermometry, such as time-domain thermoreflectance\cite{Ziabari.2018}, Raman spectroscopy\cite{Balandin.2008, Ghosh.2008}, scanning thermal probe microscopy\cite{Majumdar.1999,Kim.2012,Menges.2016}, polymer imprint thermal mapping\cite{Kinkhabwala.2016} and electron energy loss spectroscopy\cite{Mecklenburg.2015}, providing means to locally map the sample temperature.\\

\begin{figure} 
  \includegraphics[width=\linewidth]{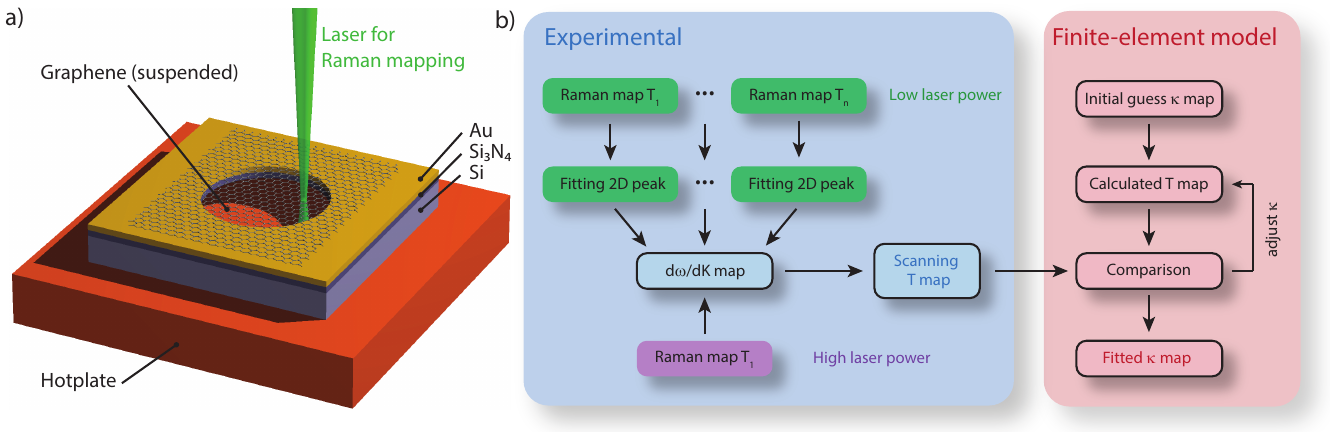}
  \caption{Experimental and Finite Element Method description. a) Schematic drawing of the suspended graphene membrane. b) Experimental workflow to obtain a temperature map upon laser illumination and computational workflow to fit the corresponding thermal conductivity map.}
  \label{fig:intro}
\end{figure}

Here, we introduce an opto-thermal method that allows for two-dimensional mapping of thermal conductivity of suspended graphene membranes. The presented method relies on a combination of scanning $\mu$-Raman spectroscopy with finite-element method (FEM) calculations. The workflow for our approach is presented in Figure.~\ref{fig:intro}. In the first experimental stage, a series of two-dimensional Raman spectroscopy maps are used to construct a temperature map of the membrane upon illumination. More specifically, Raman maps are recorded for various hotplate temperatures and at a low laser power. This series of maps is used to construct a calibration map of the Raman peak shifts with temperature. Then, another Raman map is recorded at high-laser-power, which, combined with the calibration map, is used to construct a temperature map of the membrane upon laser heating. 

The constructed experimental temperature map is then used as an input for the FEM-based fit procedure. For a given initial guess of the thermal conductivity, the lattice temperature upon laser illumination is calculated. The thermal conductivity is then iteratively adjusted, until the computed temperature map matches the experimental one. We apply this fit procedure to extract the thermal conductivity of a pristine graphene membrane that is suspended over a silicon nitride frame. Finally, we demonstrate that the thermal conductivity of the graphene membrane can be tuned in a controlled way by the introduction of helium-ion (He$^{+}$-ion) induced defects in the membrane.

\section{Results}

\subsection{Experimental temperature maps}

The CVD-graphene membranes are prepared as described in section 1 of the Supporting Information and previously. \cite{Celebi.2014,Thodkar.2016,Buchheim.2016,Braun.2021}. We applied Raman spectroscopy to obtain the lattice temperature of the suspended graphene membranes (for details see section 2 of the Supporting Information). Here, we focus on the 2D-band due to its high sensitivity to temperature changes\cite{Lee.2017,Chen.2011} of around -0.07~cm$^{-1}$/K. Alternatively, one can also rely on the G-peak due to its high linearity in peak shift versus temperature.\cite{Ghosh.2008,Cai.2010,Calizo.2007}. 

\begin{figure}
\includegraphics[width=\linewidth]{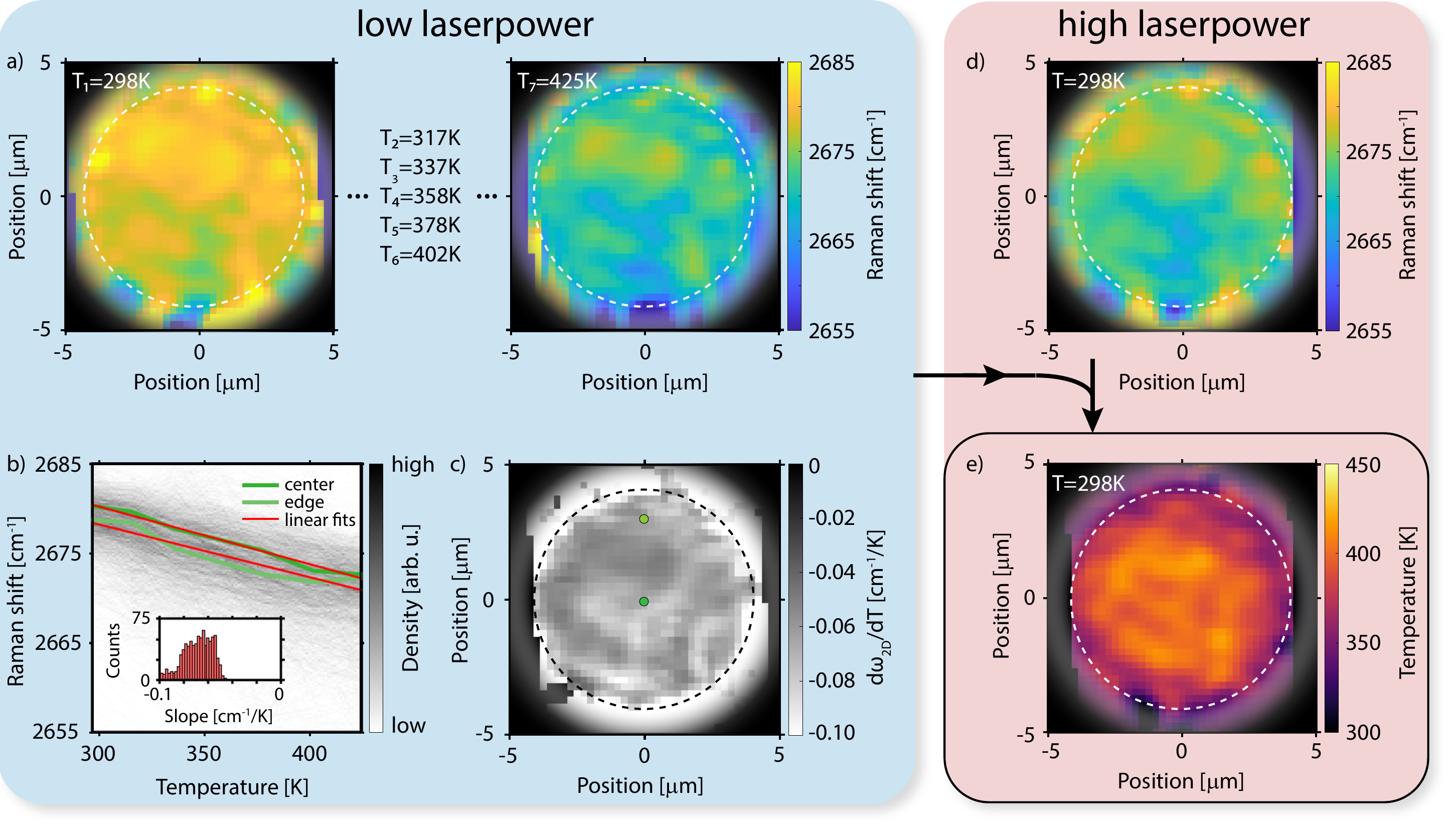}
\caption{Experimental determination of temperature map. Spatially resolved mapping of laser-induced temperature rise of graphene. a) Raman 2D-peak position obtained with $P_{laser}$ = 0.25~mW at different hot plate temperatures T$_{1}$ and T$_{7}$. The dashed circle is a guide to the eye for the support edge. b) Density plot of the temperature evolution of the 2D-peak frequency. Two spatial points (center and edge, see dots in panel c)) are highlighted to represent the method. The inset shows a histogram of $d\omega_{2D}/dT$ of the complete membrane. c) Spatial distribution of change in Raman shift per temperature change $d\omega_{2D}/dT$ obtained from linear fits to the data shown in b). d) Raman 2D-peak position obtained with $P_{laser}$ = 4~mW at 297~K. e) Temperature distribution obtained by combining the results from c) and d).}
\label{fig:Calibration}
\end{figure}

Figure~\ref{fig:Calibration}a) presents maps of the Lorentzian-fitted 2D-peaks acquired at temperatures T$_1$-T$_7$ ranging between 298~K and 425~K. The Raman spectra have been acquired at low laser power (0.25~mW) to limit any heating effects using a 532~nm excitation laser (for details see section 4 and 5 of the Supporting Information). For each pixel, the peak shift with hot plate temperature ($d\omega_{2D}/dT$) is fitted using a first-order polynomial, as shown in Figure~\ref{fig:Calibration}b). The inset presents a histogram of the slopes, showing a Gaussian distribution centered around -0.07~cm$^{-1}$/K. The spatial distribution of the peak shifts with temperature is displayed in Figure~\ref{fig:Calibration}c), showing substantial spatial variations. 
Once this calibration map is acquired, a Raman map of the graphene membrane is acquired at high laser power (4~mW), as shown in Figure~\ref{fig:Calibration}d). The high laser power causes the graphene to locally heat, resulting in a shift in the Raman peak position. By combining this high-power measurement with the $d\omega_{2D}/dT$ map obtained at low laser power, a map of the average temperature within the laser spot is obtained for each laser position, as shown in Figure~\ref{fig:Calibration}b). We note that local variations in the temperature upon illumination on the order of 50-100~K are observed, highlighting the importance of spatially mapping the temperature, superior to other techniques that extract thermal conductivity from a single spot.

\subsection{FEM calculations}

\begin{figure} 
  \includegraphics[width=\linewidth]{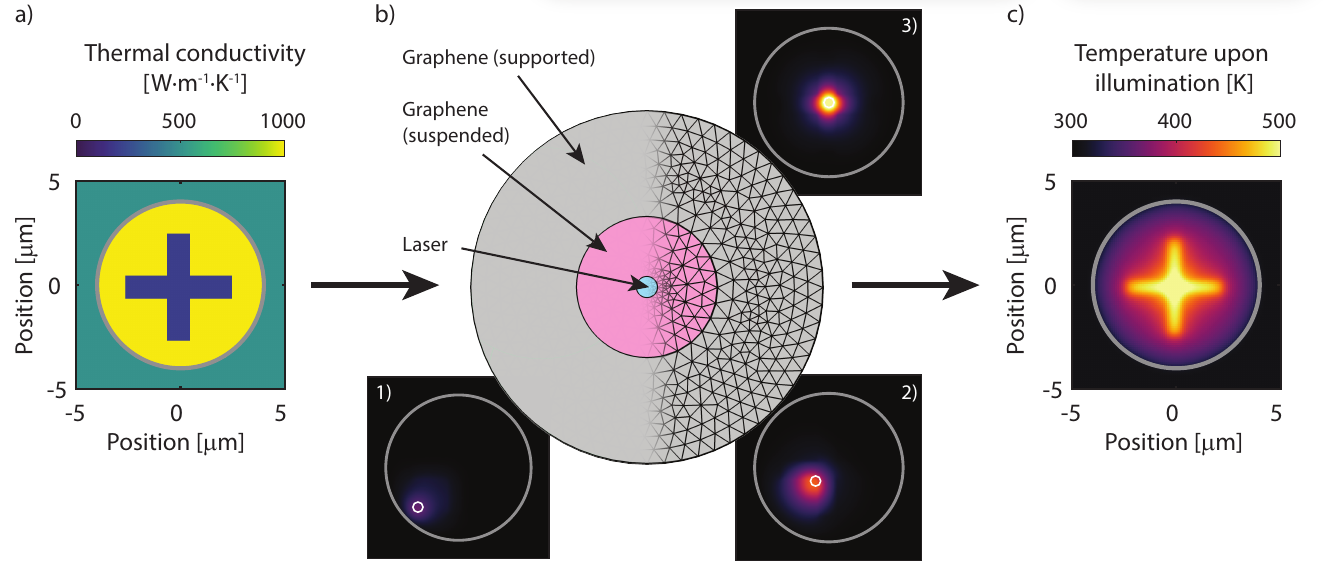}
  \caption{Finite Element Method description. a) Input thermal conductivity map. b) Schematic representation of the sample and the calculation mesh. Temperature profiles of the graphene membrane with the heating laser spot at three different positions 1)-3). c) Temperature profile upon laser illumination.}
  \label{fig:FEM}
\end{figure}

FEM calculations are employed for the computation of the temperature map of the system upon laser illumination for a given spatial thermal conductivity distribution. This concept is illustrated in Figure~\ref{fig:FEM}. As an input, a two-dimensional map of the thermal conductivity is provided, of which an example is shown Figure~\ref{fig:FEM}a). Figure~\ref{fig:FEM}b) presents the layout of the system (not to scale). A detailed description of the FEM calculation and the implementation thereof is provided in section S3 of the Supporting Information. While scanning the laser across the membrane, the full temperature distribution is calculated for each laser spot position on the membrane (three examples are provided in Figure~\ref{fig:FEM}b).\\
For each of these temperature distributions, the average temperature within the laser spot is calculated, from which all values are combined to obtain a two-dimensional map of the graphene temperature upon illumination. This induced-temperature map is presented in Figure~\ref{fig:FEM}c) and represents the same  temperature map that is obtained experimentally upon illumination of the sample with high laser power.\\

To obtain the thermal conductivity map, an iterative minimization procedure is employed. More information can be found in section S3 of the Supporting Information, in which we also validate the numerical method on a simulated system with a known thermal conductivity map. The starting point is an initial (typically uniform) guess of the thermal conductivity. In each iteration of the process, the corresponding induced temperature map is compared to the experimental temperature map, after which the thermal conductivity is adjusted pixel-wise according to the temperature difference. This process is repeated until convergence is reached. 

\subsection{Thermal conductivity map based on fitted experimental temperature map}
The induced temperature map obtained in Figure~\ref{fig:Calibration} is used as input for the iterative procedure to obtain the thermal conductivity map. Here, as clarified in section 4 of the Supporting Information, we use a uniform absorption of 2.7~\% for the suspended graphene and double that value (5.4~\%) for the supported graphene. Moreover, all the employed model parameters are summarized in Table S1 of the SI. Importantly, in the model, a diffraction-limited spot size of 300~nm is assumed. Moreover, the thermal conductivity of graphene on the supporting part, the thermal coupling to the substrate, as well as the convection parameter are taken from literature\cite{Cai.2010}.
Finally, we note that it is challenging to model the transition from the suspended graphene to the supported graphene once the laser spot is in the vicinity of the edge due to 1.) reflection from of the laser excitation at the edges of the support may lead to an increase in the deposited laser power, 2.) quenching of the Raman scattered light on the substrate may lead to an overestimation of the local temperature as the 2D peak of the suspended graphene is more pronounced than that of the supported graphene.  
To circumvent this issue, the thermal conductivity of the first 0.5~$\mu m$ of the membranes away from the support are not fitted and kept at a fixed value. Finally, the thermal conductivity is fitted for 100 iterations, after which the absorption is fitted for the same number of cycles. More details about this procedure can be found in section 2 of the SI. For numerical stability reasons, we put a lower value on the thermal conductivity at 100~Wm$^{-1}$K$^{-1}$. \\

\begin{figure}
\includegraphics[width=\linewidth]{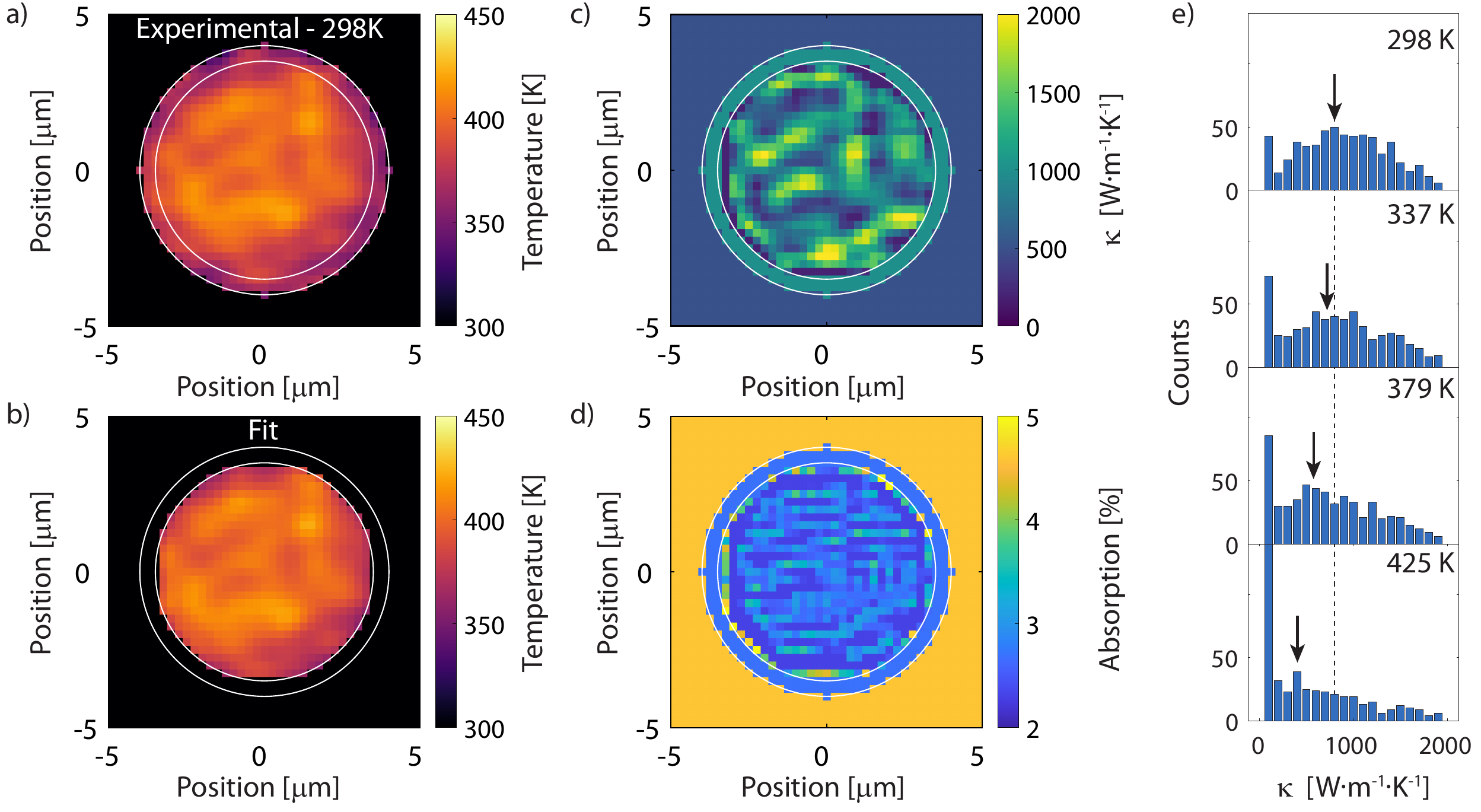}
\caption{Thermal conductivity map based on fitted experimental data. a) Experimentally determined temperature map. b) Fitted temperature map. c) Fitted thermal conductivity map. d) Converged absorption map. e) Histogram of the thermal conductivity for various hot plate temperatures. Arrows indicate the mode of the smoothed distributions.
}
\label{fig:fits_pre}
\end{figure}

Figure~\ref{fig:fits_pre}a) presents the experimental temperature map, as obtained in Figure~\ref{fig:Calibration}e), alongside the fitted temperature map in Figure~\ref{fig:fits_pre}b. The two maps closely resemble each other. The corresponding thermal conductivity map is presented in Figure~\ref{fig:fits_pre}c). We find that the local thermal conductivity values range from 500 to 2000~Wm$^{-1}$K$^{-1}$ with an average value of 1007$\pm$450~Wm$^{-1}$K$^{-1}$ highlighting the importance of spatially resolving the thermal conductivity. The average value is in agreement with values reported previously as well as by other methods.\cite{Pettes.2011,Xu.2014} In Figure~\ref{fig:fits_pre}e), we present a histogram of the fitted thermal conductivity map for increasing hot plate temperatures. The bar plots show that for increasing temperature  a gradual decrease in thermal conductivity is observed. This behavior follows the trend observed by others \cite{Chen.2011, Chen.2012, Lee.2017}.

\subsection{Thermal conductivity of defect-engineered graphene}

As a final demonstration of the capability of the presented method, we study the thermal conductivity of a graphene membrane that is exposed to He$^{+}$-ions using focused ion beam lithography. As shown previously, He$^{+}$-ions can be used to induce, in a controlled fashion, defects in suspended graphene membranes and other two-dimensional materials\cite{Buchheim.2016,Iberi.2016}.

\begin{figure}
\includegraphics[width=\linewidth]{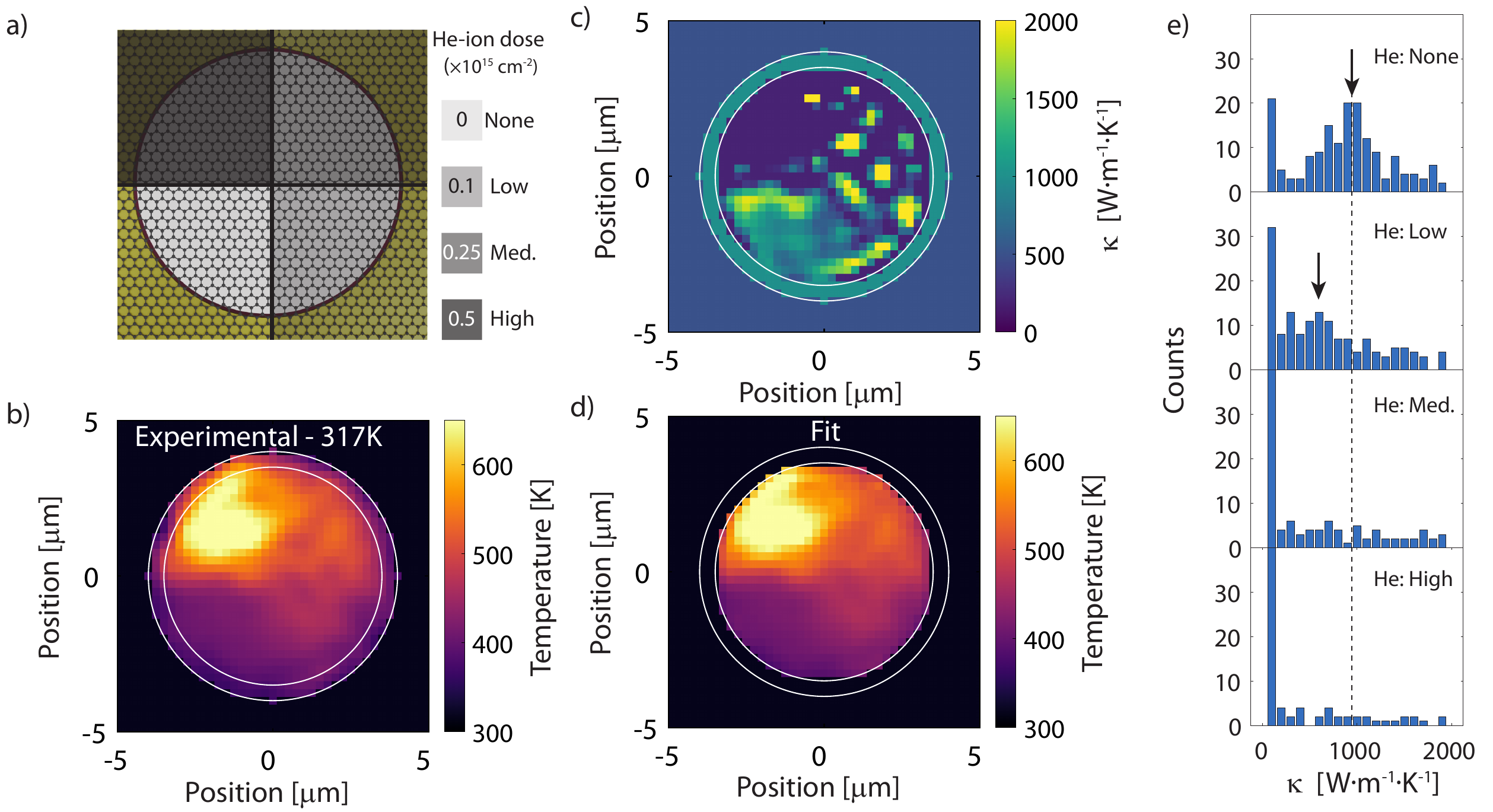}
\caption{Thermal conductivity of defect engineered graphene. a) Schematic image of a suspended graphene membrane. The areas where the membrane was exposed to He$^{+}$-ions and their corresponding dose is indicated with different colors. b) Experimentally observed temperature map. c) Fitted thermal conductivity map. d) Fitted temperature map. e) Histogram of the thermal conductivity for various defect densities. Arrows indicate the mode of the smoothed distributions where relevant.}
\label{fig:fits_post}
\end{figure}

Figure~\ref{fig:fits_post}a) presents the exposure pattern as well as the used irradiation doses. The membrane is divided into four quadrants, with the He$^{+}$-ion irradiation steadily increasing in the counter-clockwise direction, starting in the lower left with no He$^{+}$-ion dose. Manually selected representative Raman spectra of each quadrant are presented in Figure~S10 of the Supporting Information, exhibiting all the characteristic graphene peaks. The selection of the representative Raman spectra can also be done by an advanced clustering approach to avoid any human bias as described elsewhere.\cite{ElAbbassi.2021} Upon an increase of the He$^{+}$-ion dose, the D-band intensity steadily increases. The intensity ratio of the D and D' band I(D)/I(D') upon irradiation, is indicative of the type of defect.\cite{Eckmann.2012}. We extract this ratio by fitting the ratios I(D)/I(G) versus I(D')/I(G) for various He$^{+}$-ion doses. We find an intensity ratio of $\sim$11.7 for the defect-engineered graphene. This value is comparable to the reported intensity ratio of $\sim$13 for sp$^3$ type of defects (see section 6 of the Supporting Information).
We employ the same procedure as presented in Figures~\ref{fig:Calibration} for the extraction of the temperature. Figure~\ref{fig:fits_post}b) presents the induced temperature map upon a 4~mW laser illumination. In this plot, the four quadrants are visible, with the lowest temperatures recorded in the (unexposed) lower left section of the membrane, and the highest one in the upper left (most exposed). This temperature map is used as input for the iterative FEM-based fitting procedure, resulting in the fitted thermal conductivity map in Figure~\ref{fig:fits_post}c) and the corresponding fitted temperature map shown in Figure~\ref{fig:fits_post}d). Figure~\ref{fig:fits_post}e) presents a histogram of the thermal conductivity of each of the four quadrants. A steady decrease in average conductivity is observed, from $\sim$1000~Wm$^{-1}$K$^{-1}$ for the no He$^{+}$-ion irradiation, and $\sim$100~Wm$^{-1}$K$^{-1}$ for the highest He$^{+}$-ion dose. This decrease in thermal conductivity with increasing defect density is in agreement with previous reports\cite{Zhao.2015, Malekpour.2016}.

\section{Discussion}
The FEM calculations employed here assume that the heat transport through the system is following Fourier's Law. This assumption implies that the phonon mean free path is much smaller than the membrane size. Indeed, ballistic phonon transport has been reported in several nanosystems at room temperature: In substrate-supported graphene the phonon mean free path is $\sim$100~nm\cite{Bae.2013}; for suspended graphene discs, the transition from ballistic to diffusive transport occurs at $\sim$775~nm\cite{ElSachat.2019} while in ultra-thin nanowires phonon mean free paths of several micrometers have been observed\cite{Vakulov.2020}.  
Therefore, the resolution of the presented method is limited by the phonon mean free path as a spatial mapping of the thermal conductivity below this length scale would require a heat transport description based on the Boltzmann transport equation\cite{Fugallo.2014,Cepellotti.2015,Simoncelli.2020}. Given this boundary condition, the resolution of 250~nm used in this study is close to the ultimate resolution this FEM-based method allows.

A limitation of the presented method is the time consumption of the temperature calibration, reducing its use in high-throughput applications. As the laser power is low, acquiring the two-dimensional Raman map at each temperature requires several hours. This long acquisition time can also lead to a drift in the sample position during the measurement. To reduce this drift, clamping of the sample and a good thermalization of the sample with the environment is crucial. Furthermore, as changing the hot plate temperature leads to shifts of the sample position, the Raman maps acquired at various temperatures need to be aligned one versus the other. 

A second limitation is the fixed value for the absorption of 2.7~\% that is used for the first 100 cycles of the fitting procedure, after which the absorption is fitted as well. The accuracy of the model may be improved by experimentally determining the absorption at the various hot plate temperatures. Ideally, the absorption would be measured by simultaneously monitoring the transmitted, and reflected laser power while scanning across the sample. We stress that simultaneously measuring both components is crucial, as contaminations and residues on the membrane may scatter the laser light, leading to a reduction in the transmitted light, but not to an increase in absorption. However, such a measurement is challenging and technically unfeasible in our current setup.

Despite the previously mentioned limitations, our method is well suited for studying the thermal properties of two-dimensional materials, in particular for materials with an anisotropic thermal conductivity\cite{Luo.2015,Kang.2017,Islam.2018}. The method can also be extended to characterize van der Waals materials consisting of multiple layers. 
Furthermore, the method is extensible from two to three dimensions, allowing for modeling of more complex device geometries, including, for instance, stacks of 2D-materials, or the presence of contact electrodes of finite thickness.
As such, it could be used for assessing the material quality after device integration. Also, as the individual two-dimensional materials in a stacked geometry each have a distinct Raman signature, it is possible to investigate the subsurface thermal properties of materials, such as, for instance, graphene embedded in a thin hexagonal boron nitride layer. Alternatively, when the material under study is on a substrate or thick enough, other means of determining the temperature map may be used, like time-domain thermoreflectance, for reduced measurement time and improve throughput.

\section{Conclusion}

We have introduced a method for spatially mapping the thermal conductivity of single-layer graphene using a combination of Raman spectroscopy and finite-element calculations at ultimate resolution. We anticipate that this method can be applied to other single- and few layer materials. We applied the method to obtain the thermal conductivity of a pristine and He$^{+}$-ion patterned suspended single-layer graphene film. For the unpatterned film, large variations of the extracted thermal conductivity are observed and attributed to local irregularities such as contamination, defects, or folds. These findings highlight the importance of spatial mapping of the thermal conductivity, in contrast to measurement approaches that yield a thermal conductivity averaged across the entire sample. On the patterned membrane, we demonstrate controlled engineering of the thermal conductivity by He$^{+}$-ion irradiation.
As Raman spectroscopy is widely used in the two-dimensional materials community, our method is ideally suited for studying the thermal properties of other layered materials. Moreover, the working principle of the FEM method can easily be extended to more complex geometries or interfaces, in particular combined with alternative measurement techniques for providing a temperature map. Our method enables spatially resolving the thermal conductivity of atomically thin materials, a prerequisite for optimizing and engineering thermal stewardship in nanoscale devices.

\begin{acknowledgement}

This work was supported by the EC H2020 FET Open project no. 767187 (QuIET).
M.L.P. acknowledges funding by the EMPAPOSTDOCS-II program, which has received funding from the European Union’s Horizon 2020 research and innovation program under the Marie Skłodowska–Curie Grant Agreement no. 754364. M.L.P. also acknowledges funding from the Swiss National Science Foundation under the Spark grant no. 196795.
I.Z. and M.C. acknowledge funding from the Swiss National Science Foundation under the Sinergia grant no. 189924 (Hydronics). I.Z. acknowledges funding form the European Research Council (ERC) under the European Union's Horizon 2020 research and innovation program (Grant Agreement 756365). The author acknowledge support from the Multiphysics Hub @ Empa for the COMSOL Multiphyics calculations.
We thank the Cleanroom Operations Team of the Binnig and Rohrer Nanotechnology Center (BRNC) for their help and support, and Roman M. Wyss for fruitful discussions and supply of Si$_{3}$N$_{4}$ frames. We further thank Jan Overbeck, Maria El Abbassi, Marta De Luca and Milo Y. Swinkels for fruitful discussions.

\end{acknowledgement}

\section{Author contributions statement}

O.B., I.S., M.C., and M.P. conceived and designed the experiments. K.T. developed the graphene growth recipe and transfer process. R.F. performed the graphene growth. O.B and I.S. prepared the SiN frame and performed the defect engineering using a focused-ion beam. O.B. performed the Raman measurements. O.B., M.P., M.C., and I.Z. did the Raman spectroscopy analysis. P.B. developed the finite-element model to calculate the temperature distribution for a single laser spot position. M.P. extended the model to construct the temperature map upon illumination by the Raman laser and developed the procedure to fit the thermal conductivity. M.P. performed all finite-element calculations. 
O.B., M.P., and M.C. wrote the manuscript. All authors discussed the results and implications and commented on the manuscript.


\section{Competing interests}
The authors declare no competing financial interests.


\begin{suppinfo}
Available free of charge are: All codes and data used in this study are  at https://github.com/Mickael Perrin74/ThermalconductivityMapping and further Supporting Information at https://pubs.acs.org/doi/XXX/acs.nanolett.XXX., containing:\\
\begin{itemize}
    \item Sample fabrication
    \item Raman setup and spectra analysis
    \item Finite-element calculations
    \item Absorption and Laser power
    \item Temperature calibration
    \item He$^{+}$-ion irradiation
\end{itemize}

\end{suppinfo}

\bibliography{References_20210315}

\providecommand{\latin}[1]{#1}
\makeatletter
\providecommand{\doi}
  {\begingroup\let\do\@makeother\dospecials
  \catcode`\{=1 \catcode`\}=2 \doi@aux}
\providecommand{\doi@aux}[1]{\endgroup\texttt{#1}}
\makeatother
\providecommand*\mcitethebibliography{\thebibliography}
\csname @ifundefined\endcsname{endmcitethebibliography}
  {\let\endmcitethebibliography\endthebibliography}{}
\begin{mcitethebibliography}{52}
\providecommand*\natexlab[1]{#1}
\providecommand*\mciteSetBstSublistMode[1]{}
\providecommand*\mciteSetBstMaxWidthForm[2]{}
\providecommand*\mciteBstWouldAddEndPuncttrue
  {\def\EndOfBibitem{\unskip.}}
\providecommand*\mciteBstWouldAddEndPunctfalse
  {\let\EndOfBibitem\relax}
\providecommand*\mciteSetBstMidEndSepPunct[3]{}
\providecommand*\mciteSetBstSublistLabelBeginEnd[3]{}
\providecommand*\EndOfBibitem{}
\mciteSetBstSublistMode{f}
\mciteSetBstMaxWidthForm{subitem}{(\alph{mcitesubitemcount})}
\mciteSetBstSublistLabelBeginEnd
  {\mcitemaxwidthsubitemform\space}
  {\relax}
  {\relax}

\bibitem[Shi \latin{et~al.}(2015)Shi, Dames, Lukes, Reddy, Duda, Cahill, Lee,
  Marconnet, Goodson, Bahk, Shakouri, Prasher, Felts, King, Han, and
  Bischof]{Shi.2015}
Shi,~L. \latin{et~al.}  Evaluating Broader Impacts of Nanoscale Thermal
  Transport Research. \emph{Nanoscale and Microscale Thermophysical
  Engineering} \textbf{2015}, \emph{19}, 127--165\relax
\mciteBstWouldAddEndPuncttrue
\mciteSetBstMidEndSepPunct{\mcitedefaultmidpunct}
{\mcitedefaultendpunct}{\mcitedefaultseppunct}\relax
\EndOfBibitem
\bibitem[Song \latin{et~al.}(2018)Song, Liu, Liu, Wu, Cheng, and
  Kang]{Song.2018}
Song,~H.; Liu,~J.; Liu,~B.; Wu,~J.; Cheng,~H.-M.; Kang,~F. Two-Dimensional
  Materials for Thermal Management Applications. \emph{Joule} \textbf{2018},
  \emph{2}, 442--463\relax
\mciteBstWouldAddEndPuncttrue
\mciteSetBstMidEndSepPunct{\mcitedefaultmidpunct}
{\mcitedefaultendpunct}{\mcitedefaultseppunct}\relax
\EndOfBibitem
\bibitem[Corbino(1910)]{Corbino.1910}
Corbino,~O.~M. Thermal oscillations in lamps of thin fibers with alternating
  current flowing through them and the resulting effect on the rectifier as a
  result of the presence of even-numbered harmonics. \emph{Physikalische
  Zeitschrift} \textbf{1910}, \emph{11}, 413--417\relax
\mciteBstWouldAddEndPuncttrue
\mciteSetBstMidEndSepPunct{\mcitedefaultmidpunct}
{\mcitedefaultendpunct}{\mcitedefaultseppunct}\relax
\EndOfBibitem
\bibitem[Corbino(1911)]{Corbino.1911}
Corbino,~O.~M. Periodic resistance changes of fine metal threads which are
  brought together by alternating streams as well as deduction of their thermo
  characteristics at high temperatures. \emph{Physikalische Zeitschrift}
  \textbf{1911}, \emph{12}, 292--295\relax
\mciteBstWouldAddEndPuncttrue
\mciteSetBstMidEndSepPunct{\mcitedefaultmidpunct}
{\mcitedefaultendpunct}{\mcitedefaultseppunct}\relax
\EndOfBibitem
\bibitem[Seol \latin{et~al.}(2010)Seol, Jo, Moore, Lindsay, Aitken, Pettes, Li,
  Yao, Huang, Broido, Mingo, Ruoff, and Shi]{Seol.2010}
Seol,~J.~H.; Jo,~I.; Moore,~A.~L.; Lindsay,~L.; Aitken,~Z.~H.; Pettes,~M.~T.;
  Li,~X.; Yao,~Z.; Huang,~R.; Broido,~D.; Mingo,~N.; Ruoff,~R.~S.; Shi,~L.
  Two-Dimensional Phonon Transport in Supported Graphene. \emph{Science}
  \textbf{2010}, \emph{328}, 213--216\relax
\mciteBstWouldAddEndPuncttrue
\mciteSetBstMidEndSepPunct{\mcitedefaultmidpunct}
{\mcitedefaultendpunct}{\mcitedefaultseppunct}\relax
\EndOfBibitem
\bibitem[Swinkels \latin{et~al.}(2015)Swinkels, {van Delft}, Oliveira, Cavalli,
  Zardo, {van der Heijden}, and Bakkers]{Swinkels.2015}
Swinkels,~M.~Y.; {van Delft},~M.~R.; Oliveira,~D.~S.; Cavalli,~A.; Zardo,~I.;
  {van der Heijden},~R.~W.; Bakkers,~E. P. A.~M. Diameter dependence of the
  thermal conductivity of InAs nanowires. \emph{Nanotechnology} \textbf{2015},
  \emph{26}, 385401\relax
\mciteBstWouldAddEndPuncttrue
\mciteSetBstMidEndSepPunct{\mcitedefaultmidpunct}
{\mcitedefaultendpunct}{\mcitedefaultseppunct}\relax
\EndOfBibitem
\bibitem[Yazji \latin{et~al.}(2016)Yazji, Swinkels, de~Luca, Hoffmann,
  Ercolani, Roddaro, Abstreiter, Sorba, Bakkers, and Zardo]{Yazji.2016}
Yazji,~S.; Swinkels,~M.~Y.; de~Luca,~M.; Hoffmann,~E.~A.; Ercolani,~D.;
  Roddaro,~S.; Abstreiter,~G.; Sorba,~L.; Bakkers,~E. P. A.~M.; Zardo,~I.
  Assessing the thermoelectric properties of single InSb nanowires: the role of
  thermal contact resistance. \emph{Semiconductor Science and Technology}
  \textbf{2016}, \emph{31}, 064001\relax
\mciteBstWouldAddEndPuncttrue
\mciteSetBstMidEndSepPunct{\mcitedefaultmidpunct}
{\mcitedefaultendpunct}{\mcitedefaultseppunct}\relax
\EndOfBibitem
\bibitem[Deshpande \latin{et~al.}(2009)Deshpande, Hsieh, Bushmaker, Bockrath,
  and Cronin]{Deshpande.2009}
Deshpande,~V.~V.; Hsieh,~S.; Bushmaker,~A.~W.; Bockrath,~M.; Cronin,~S.~B.
  Spatially resolved temperature measurements of electrically heated carbon
  nanotubes. \emph{Physical review letters} \textbf{2009}, \emph{102},
  105501\relax
\mciteBstWouldAddEndPuncttrue
\mciteSetBstMidEndSepPunct{\mcitedefaultmidpunct}
{\mcitedefaultendpunct}{\mcitedefaultseppunct}\relax
\EndOfBibitem
\bibitem[Soini \latin{et~al.}(2010)Soini, Zardo, Uccelli, Funk, Koblm{\"u}ller,
  {Fontcuberta i Morral}, and Abstreiter]{Soini.2010}
Soini,~M.; Zardo,~I.; Uccelli,~E.; Funk,~S.; Koblm{\"u}ller,~G.; {Fontcuberta i
  Morral},~A.; Abstreiter,~G. Thermal conductivity of GaAs nanowires studied by
  micro-Raman spectroscopy combined with laser heating. \emph{Applied Physics
  Letters} \textbf{2010}, \emph{97}, 263107\relax
\mciteBstWouldAddEndPuncttrue
\mciteSetBstMidEndSepPunct{\mcitedefaultmidpunct}
{\mcitedefaultendpunct}{\mcitedefaultseppunct}\relax
\EndOfBibitem
\bibitem[Reparaz \latin{et~al.}(2014)Reparaz, Chavez-Angel, Wagner,
  Graczykowski, Gomis-Bresco, Alzina, and {Sotomayor Torres}]{Reparaz.2014}
Reparaz,~J.~S.; Chavez-Angel,~E.; Wagner,~M.~R.; Graczykowski,~B.;
  Gomis-Bresco,~J.; Alzina,~F.; {Sotomayor Torres},~C.~M. A novel contactless
  technique for thermal field mapping and thermal conductivity determination:
  two-laser Raman thermometry. \emph{The Review of scientific instruments}
  \textbf{2014}, \emph{85}, 034901\relax
\mciteBstWouldAddEndPuncttrue
\mciteSetBstMidEndSepPunct{\mcitedefaultmidpunct}
{\mcitedefaultendpunct}{\mcitedefaultseppunct}\relax
\EndOfBibitem
\bibitem[Neogi \latin{et~al.}(2015)Neogi, Reparaz, Pereira, Graczykowski,
  Wagner, Sledzinska, Shchepetov, Prunnila, Ahopelto, Sotomayor-Torres, and
  Donadio]{Neogi.2015}
Neogi,~S.; Reparaz,~J.~S.; Pereira,~L. F.~C.; Graczykowski,~B.; Wagner,~M.~R.;
  Sledzinska,~M.; Shchepetov,~A.; Prunnila,~M.; Ahopelto,~J.;
  Sotomayor-Torres,~C.~M.; Donadio,~D. Tuning thermal transport in ultrathin
  silicon membranes by surface nanoscale engineering. \emph{ACS nano}
  \textbf{2015}, \emph{9}, 3820--3828\relax
\mciteBstWouldAddEndPuncttrue
\mciteSetBstMidEndSepPunct{\mcitedefaultmidpunct}
{\mcitedefaultendpunct}{\mcitedefaultseppunct}\relax
\EndOfBibitem
\bibitem[Shahil and Balandin(2012)Shahil, and Balandin]{Shahil.2012}
Shahil,~K.~M.; Balandin,~A.~A. Thermal properties of graphene and multilayer
  graphene: Applications in thermal interface materials. \emph{Solid State
  Communications} \textbf{2012}, \emph{152}, 1331--1340\relax
\mciteBstWouldAddEndPuncttrue
\mciteSetBstMidEndSepPunct{\mcitedefaultmidpunct}
{\mcitedefaultendpunct}{\mcitedefaultseppunct}\relax
\EndOfBibitem
\bibitem[Balandin(2020)]{Balandin.2020}
Balandin,~A.~A. Phononics of Graphene and Related Materials. \emph{ACS nano}
  \textbf{2020}, \relax
\mciteBstWouldAddEndPunctfalse
\mciteSetBstMidEndSepPunct{\mcitedefaultmidpunct}
{}{\mcitedefaultseppunct}\relax
\EndOfBibitem
\bibitem[Kasirga(2020)]{Kasirga.2020}
Kasirga,~T.~S. \emph{Thermal Conductivity Measurements in Atomically Thin
  Materials and Devices}; {Springer Nature}, 2020\relax
\mciteBstWouldAddEndPuncttrue
\mciteSetBstMidEndSepPunct{\mcitedefaultmidpunct}
{\mcitedefaultendpunct}{\mcitedefaultseppunct}\relax
\EndOfBibitem
\bibitem[Balandin \latin{et~al.}(2008)Balandin, Ghosh, Bao, Calizo,
  Teweldebrhan, Miao, and Lau]{Balandin.2008}
Balandin,~A.~A.; Ghosh,~S.; Bao,~W.; Calizo,~I.; Teweldebrhan,~D.; Miao,~F.;
  Lau,~C.~N. Superior thermal conductivity of single-layer graphene. \emph{Nano
  letters} \textbf{2008}, \emph{8}, 902--907\relax
\mciteBstWouldAddEndPuncttrue
\mciteSetBstMidEndSepPunct{\mcitedefaultmidpunct}
{\mcitedefaultendpunct}{\mcitedefaultseppunct}\relax
\EndOfBibitem
\bibitem[Ghosh \latin{et~al.}(2008)Ghosh, Calizo, Teweldebrhan, Pokatilov,
  Nika, Balandin, Bao, Miao, and Lau]{Ghosh.2008}
Ghosh,~S.; Calizo,~I.; Teweldebrhan,~D.; Pokatilov,~E.~P.; Nika,~D.~L.;
  Balandin,~A.~A.; Bao,~W.; Miao,~F.; Lau,~C.~N. Extremely high thermal
  conductivity of graphene: Prospects for thermal management applications in
  nanoelectronic circuits. \emph{Applied Physics Letters} \textbf{2008},
  \emph{92}, 151911\relax
\mciteBstWouldAddEndPuncttrue
\mciteSetBstMidEndSepPunct{\mcitedefaultmidpunct}
{\mcitedefaultendpunct}{\mcitedefaultseppunct}\relax
\EndOfBibitem
\bibitem[Faugeras \latin{et~al.}(2010)Faugeras, Faugeras, Orlita, Potemski,
  Nair, and Geim]{Faugeras.2010}
Faugeras,~C.; Faugeras,~B.; Orlita,~M.; Potemski,~M.; Nair,~R.~R.; Geim,~A.~K.
  Thermal conductivity of graphene in corbino membrane geometry. \emph{ACS
  nano} \textbf{2010}, \emph{4}, 1889--1892\relax
\mciteBstWouldAddEndPuncttrue
\mciteSetBstMidEndSepPunct{\mcitedefaultmidpunct}
{\mcitedefaultendpunct}{\mcitedefaultseppunct}\relax
\EndOfBibitem
\bibitem[Lee \latin{et~al.}(2011)Lee, Yoon, Kim, Lee, and Cheong]{Lee.2011}
Lee,~J.-U.; Yoon,~D.; Kim,~H.; Lee,~S.~W.; Cheong,~H. Thermal conductivity of
  suspended pristine graphene measured by Raman spectroscopy. \emph{Physical
  Review B} \textbf{2011}, \emph{83}\relax
\mciteBstWouldAddEndPuncttrue
\mciteSetBstMidEndSepPunct{\mcitedefaultmidpunct}
{\mcitedefaultendpunct}{\mcitedefaultseppunct}\relax
\EndOfBibitem
\bibitem[Cai \latin{et~al.}(2010)Cai, Moore, Zhu, Li, Chen, Shi, and
  Ruoff]{Cai.2010}
Cai,~W.; Moore,~A.~L.; Zhu,~Y.; Li,~X.; Chen,~S.; Shi,~L.; Ruoff,~R.~S. Thermal
  transport in suspended and supported monolayer graphene grown by chemical
  vapor deposition. \emph{Nano letters} \textbf{2010}, \emph{10},
  1645--1651\relax
\mciteBstWouldAddEndPuncttrue
\mciteSetBstMidEndSepPunct{\mcitedefaultmidpunct}
{\mcitedefaultendpunct}{\mcitedefaultseppunct}\relax
\EndOfBibitem
\bibitem[Chen \latin{et~al.}(2011)Chen, Moore, Cai, Suk, An, Mishra, Amos,
  Magnuson, Kang, Shi, and Ruoff]{Chen.2011}
Chen,~S.; Moore,~A.~L.; Cai,~W.; Suk,~J.~W.; An,~J.; Mishra,~C.; Amos,~C.;
  Magnuson,~C.~W.; Kang,~J.; Shi,~L.; Ruoff,~R.~S. Raman measurements of
  thermal transport in suspended monolayer graphene of variable sizes in vacuum
  and gaseous environments. \emph{ACS nano} \textbf{2011}, \emph{5},
  321--328\relax
\mciteBstWouldAddEndPuncttrue
\mciteSetBstMidEndSepPunct{\mcitedefaultmidpunct}
{\mcitedefaultendpunct}{\mcitedefaultseppunct}\relax
\EndOfBibitem
\bibitem[Chen \latin{et~al.}(2012)Chen, Wu, Mishra, Kang, Zhang, Cho, Cai,
  Balandin, and Ruoff]{Chen.2012}
Chen,~S.; Wu,~Q.; Mishra,~C.; Kang,~J.; Zhang,~H.; Cho,~K.; Cai,~W.;
  Balandin,~A.~A.; Ruoff,~R.~S. Thermal conductivity of isotopically modified
  graphene. \emph{Nature materials} \textbf{2012}, \emph{11}, 203--207\relax
\mciteBstWouldAddEndPuncttrue
\mciteSetBstMidEndSepPunct{\mcitedefaultmidpunct}
{\mcitedefaultendpunct}{\mcitedefaultseppunct}\relax
\EndOfBibitem
\bibitem[Lee \latin{et~al.}(2017)Lee, Kihm, Kim, Shin, Lee, Park, Cheon, Kwon,
  Lim, and Lee]{Lee.2017}
Lee,~W.; Kihm,~K.~D.; Kim,~H.~G.; Shin,~S.; Lee,~C.; Park,~J.~S.; Cheon,~S.;
  Kwon,~O.~M.; Lim,~G.; Lee,~W. In-Plane Thermal Conductivity of
  Polycrystalline Chemical Vapor Deposition Graphene with Controlled Grain
  Sizes. \emph{Nano letters} \textbf{2017}, \emph{17}, 2361--2366\relax
\mciteBstWouldAddEndPuncttrue
\mciteSetBstMidEndSepPunct{\mcitedefaultmidpunct}
{\mcitedefaultendpunct}{\mcitedefaultseppunct}\relax
\EndOfBibitem
\bibitem[Ma \latin{et~al.}(2017)Ma, Liu, Wen, Gao, Ren, Chen, Jin, Ma, Xu,
  Cheng, and Ren]{Ma.2017}
Ma,~T.; Liu,~Z.; Wen,~J.; Gao,~Y.; Ren,~X.; Chen,~H.; Jin,~C.; Ma,~X.-L.;
  Xu,~N.; Cheng,~H.-M.; Ren,~W. Tailoring the thermal and electrical transport
  properties of graphene films by grain size engineering. \emph{Nature
  communications} \textbf{2017}, \emph{8}, 14486\relax
\mciteBstWouldAddEndPuncttrue
\mciteSetBstMidEndSepPunct{\mcitedefaultmidpunct}
{\mcitedefaultendpunct}{\mcitedefaultseppunct}\relax
\EndOfBibitem
\bibitem[Chen \latin{et~al.}(2012)Chen, Li, Zhang, Qu, Ji, Ruoff, and
  Cai]{Chen.2012c}
Chen,~S.; Li,~Q.; Zhang,~Q.; Qu,~Y.; Ji,~H.; Ruoff,~R.~S.; Cai,~W. Thermal
  conductivity measurements of suspended graphene with and without wrinkles by
  micro-Raman mapping. \emph{Nanotechnology} \textbf{2012}, \emph{23},
  365701\relax
\mciteBstWouldAddEndPuncttrue
\mciteSetBstMidEndSepPunct{\mcitedefaultmidpunct}
{\mcitedefaultendpunct}{\mcitedefaultseppunct}\relax
\EndOfBibitem
\bibitem[Zhao \latin{et~al.}(2015)Zhao, Wang, Wu, Wang, Bi, Liang, Yang, Chen,
  Xu, and Ni]{Zhao.2015}
Zhao,~W.; Wang,~Y.; Wu,~Z.; Wang,~W.; Bi,~K.; Liang,~Z.; Yang,~J.; Chen,~Y.;
  Xu,~Z.; Ni,~Z. Defect-Engineered Heat Transport in Graphene: A Route to High
  Efficient Thermal Rectification. \emph{Scientific reports} \textbf{2015},
  \emph{5}, 11962\relax
\mciteBstWouldAddEndPuncttrue
\mciteSetBstMidEndSepPunct{\mcitedefaultmidpunct}
{\mcitedefaultendpunct}{\mcitedefaultseppunct}\relax
\EndOfBibitem
\bibitem[Malekpour \latin{et~al.}(2016)Malekpour, Ramnani, Srinivasan,
  Balasubramanian, Nika, Mulchandani, Lake, and Balandin]{Malekpour.2016}
Malekpour,~H.; Ramnani,~P.; Srinivasan,~S.; Balasubramanian,~G.; Nika,~D.~L.;
  Mulchandani,~A.; Lake,~R.~K.; Balandin,~A.~A. Thermal conductivity of
  graphene with defects induced by electron beam irradiation. \emph{Nanoscale}
  \textbf{2016}, \emph{8}, 14608--14616\relax
\mciteBstWouldAddEndPuncttrue
\mciteSetBstMidEndSepPunct{\mcitedefaultmidpunct}
{\mcitedefaultendpunct}{\mcitedefaultseppunct}\relax
\EndOfBibitem
\bibitem[Ziabari \latin{et~al.}(2018)Ziabari, Torres, Vermeersch, Xuan,
  Cartoix{\`a}, Torell{\'o}, Bahk, Koh, Parsa, Ye, Alvarez, and
  Shakouri]{Ziabari.2018}
Ziabari,~A.; Torres,~P.; Vermeersch,~B.; Xuan,~Y.; Cartoix{\`a},~X.;
  Torell{\'o},~A.; Bahk,~J.-H.; Koh,~Y.~R.; Parsa,~M.; Ye,~P.~D.;
  Alvarez,~F.~X.; Shakouri,~A. Full-field thermal imaging of quasiballistic
  crosstalk reduction in nanoscale devices. \emph{Nature communications}
  \textbf{2018}, \emph{9}, 255\relax
\mciteBstWouldAddEndPuncttrue
\mciteSetBstMidEndSepPunct{\mcitedefaultmidpunct}
{\mcitedefaultendpunct}{\mcitedefaultseppunct}\relax
\EndOfBibitem
\bibitem[Majumdar(1999)]{Majumdar.1999}
Majumdar,~A. Scanning Thermal Microscopy. \emph{Annu. Rev. Mater. Sci.}
  \textbf{1999}, 505--585\relax
\mciteBstWouldAddEndPuncttrue
\mciteSetBstMidEndSepPunct{\mcitedefaultmidpunct}
{\mcitedefaultendpunct}{\mcitedefaultseppunct}\relax
\EndOfBibitem
\bibitem[Kim \latin{et~al.}(2012)Kim, Jeong, Lee, and Reddy]{Kim.2012}
Kim,~K.; Jeong,~W.; Lee,~W.; Reddy,~P. Ultra-high vacuum scanning thermal
  microscopy for nanometer resolution quantitative thermometry. \emph{ACS nano}
  \textbf{2012}, \emph{6}, 4248--4257\relax
\mciteBstWouldAddEndPuncttrue
\mciteSetBstMidEndSepPunct{\mcitedefaultmidpunct}
{\mcitedefaultendpunct}{\mcitedefaultseppunct}\relax
\EndOfBibitem
\bibitem[Menges \latin{et~al.}(2016)Menges, Mensch, Schmid, Riel, Stemmer, and
  Gotsmann]{Menges.2016}
Menges,~F.; Mensch,~P.; Schmid,~H.; Riel,~H.; Stemmer,~A.; Gotsmann,~B.
  Temperature mapping of operating nanoscale devices by scanning probe
  thermometry. \emph{Nature communications} \textbf{2016}, \emph{7},
  10874\relax
\mciteBstWouldAddEndPuncttrue
\mciteSetBstMidEndSepPunct{\mcitedefaultmidpunct}
{\mcitedefaultendpunct}{\mcitedefaultseppunct}\relax
\EndOfBibitem
\bibitem[Kinkhabwala \latin{et~al.}(2016)Kinkhabwala, Staffaroni, Suzer,
  Burgos, and Stipe]{Kinkhabwala.2016}
Kinkhabwala,~A.~A.; Staffaroni,~M.; Suzer,~O.; Burgos,~S.; Stipe,~B. Nanoscale
  Thermal Mapping of HAMR Heads Using Polymer Imprint Thermal Mapping.
  \emph{IEEE Transactions on Magnetics} \textbf{2016}, \emph{52}, 1--4\relax
\mciteBstWouldAddEndPuncttrue
\mciteSetBstMidEndSepPunct{\mcitedefaultmidpunct}
{\mcitedefaultendpunct}{\mcitedefaultseppunct}\relax
\EndOfBibitem
\bibitem[Mecklenburg \latin{et~al.}(2015)Mecklenburg, Hubbard, White, Dhall,
  Cronin, Aloni, and Regan]{Mecklenburg.2015}
Mecklenburg,~M.; Hubbard,~W.~A.; White,~E.~R.; Dhall,~R.; Cronin,~S.~B.;
  Aloni,~S.; Regan,~B.~C. Thermal measurement. Nanoscale temperature mapping in
  operating microelectronic devices. \emph{Science} \textbf{2015}, \emph{347},
  629--632\relax
\mciteBstWouldAddEndPuncttrue
\mciteSetBstMidEndSepPunct{\mcitedefaultmidpunct}
{\mcitedefaultendpunct}{\mcitedefaultseppunct}\relax
\EndOfBibitem
\bibitem[Celebi \latin{et~al.}(2014)Celebi, Buchheim, Wyss, Droudian, Gasser,
  Shorubalko, Kye, Lee, and Park]{Celebi.2014}
Celebi,~K.; Buchheim,~J.; Wyss,~R.~M.; Droudian,~A.; Gasser,~P.;
  Shorubalko,~I.; Kye,~J.-I.; Lee,~C.; Park,~H.~G. Ultimate permeation across
  atomically thin porous graphene. \emph{Science} \textbf{2014}, \emph{344},
  289--292\relax
\mciteBstWouldAddEndPuncttrue
\mciteSetBstMidEndSepPunct{\mcitedefaultmidpunct}
{\mcitedefaultendpunct}{\mcitedefaultseppunct}\relax
\EndOfBibitem
\bibitem[Thodkar \latin{et~al.}(2016)Thodkar, {El Abbassi}, L{\"u}{\"o}nd,
  Overney, Sch{\"o}nenberger, Jeanneret, and Calame]{Thodkar.2016}
Thodkar,~K.; {El Abbassi},~M.; L{\"u}{\"o}nd,~F.; Overney,~F.;
  Sch{\"o}nenberger,~C.; Jeanneret,~B.; Calame,~M. Comparative study of single
  and multi domain CVD graphene using large-area Raman mapping and electrical
  transport characterization. \emph{physica status solidi (RRL) - Rapid
  Research Letters} \textbf{2016}, \emph{10}, 807--811\relax
\mciteBstWouldAddEndPuncttrue
\mciteSetBstMidEndSepPunct{\mcitedefaultmidpunct}
{\mcitedefaultendpunct}{\mcitedefaultseppunct}\relax
\EndOfBibitem
\bibitem[Buchheim \latin{et~al.}(2016)Buchheim, Wyss, Shorubalko, and
  Park]{Buchheim.2016}
Buchheim,~J.; Wyss,~R.~M.; Shorubalko,~I.; Park,~H.~G. Understanding the
  interaction between energetic ions and freestanding graphene towards
  practical 2D perforation. \emph{Nanoscale} \textbf{2016}, \emph{8},
  8345--8354\relax
\mciteBstWouldAddEndPuncttrue
\mciteSetBstMidEndSepPunct{\mcitedefaultmidpunct}
{\mcitedefaultendpunct}{\mcitedefaultseppunct}\relax
\EndOfBibitem
\bibitem[Braun \latin{et~al.}(2021)Braun, Overbeck, {El Abbassi}, K{\"a}ser,
  Furrer, Olziersky, Flasby, {Borin Barin}, Darawish, M{\"u}llen, Ruffieux,
  Fasel, Shorubalko, Perrin, and Calame]{Braun.2021}
Braun,~O.; Overbeck,~J.; {El Abbassi},~M.; K{\"a}ser,~S.; Furrer,~R.;
  Olziersky,~A.; Flasby,~A.; {Borin Barin},~G.; Darawish,~R.; M{\"u}llen,~K.;
  Ruffieux,~P.; Fasel,~R.; Shorubalko,~I.; Perrin,~M.~L.; Calame,~M. Optimized
  Graphene Electrodes for contacting Graphene Nanoribbons. \emph{arXiv
  preprint} \textbf{2021}, \emph{arXiv:2102.13033}\relax
\mciteBstWouldAddEndPuncttrue
\mciteSetBstMidEndSepPunct{\mcitedefaultmidpunct}
{\mcitedefaultendpunct}{\mcitedefaultseppunct}\relax
\EndOfBibitem
\bibitem[Calizo \latin{et~al.}(2007)Calizo, Balandin, Bao, Miao, and
  Lau]{Calizo.2007}
Calizo,~I.; Balandin,~A.~A.; Bao,~W.; Miao,~F.; Lau,~C.~N. Temperature
  dependence of the Raman spectra of graphene and graphene multilayers.
  \emph{Nano letters} \textbf{2007}, \emph{7}, 2645--2649\relax
\mciteBstWouldAddEndPuncttrue
\mciteSetBstMidEndSepPunct{\mcitedefaultmidpunct}
{\mcitedefaultendpunct}{\mcitedefaultseppunct}\relax
\EndOfBibitem
\bibitem[Pettes \latin{et~al.}(2011)Pettes, Jo, Yao, and Shi]{Pettes.2011}
Pettes,~M.~T.; Jo,~I.; Yao,~Z.; Shi,~L. Influence of polymeric residue on the
  thermal conductivity of suspended bilayer graphene. \emph{Nano letters}
  \textbf{2011}, \emph{11}, 1195--1200\relax
\mciteBstWouldAddEndPuncttrue
\mciteSetBstMidEndSepPunct{\mcitedefaultmidpunct}
{\mcitedefaultendpunct}{\mcitedefaultseppunct}\relax
\EndOfBibitem
\bibitem[Xu \latin{et~al.}(2014)Xu, Pereira, Wang, Wu, Zhang, Zhao, Bae, {Tinh
  Bui}, Xie, Thong, Hong, Loh, Donadio, Li, and {\"O}zyilmaz]{Xu.2014}
Xu,~X.; Pereira,~L. F.~C.; Wang,~Y.; Wu,~J.; Zhang,~K.; Zhao,~X.; Bae,~S.;
  {Tinh Bui},~C.; Xie,~R.; Thong,~J. T.~L.; Hong,~B.~H.; Loh,~K.~P.;
  Donadio,~D.; Li,~B.; {\"O}zyilmaz,~B. Length-dependent thermal conductivity
  in suspended single-layer graphene. \emph{Nature communications}
  \textbf{2014}, \emph{5}, 3689\relax
\mciteBstWouldAddEndPuncttrue
\mciteSetBstMidEndSepPunct{\mcitedefaultmidpunct}
{\mcitedefaultendpunct}{\mcitedefaultseppunct}\relax
\EndOfBibitem
\bibitem[Iberi \latin{et~al.}(2016)Iberi, Liang, Ievlev, Stanford, Lin, Li,
  Mahjouri-Samani, Jesse, Sumpter, Kalinin, Joy, Xiao, Belianinov, and
  Ovchinnikova]{Iberi.2016}
Iberi,~V.; Liang,~L.; Ievlev,~A.~V.; Stanford,~M.~G.; Lin,~M.-W.; Li,~X.;
  Mahjouri-Samani,~M.; Jesse,~S.; Sumpter,~B.~G.; Kalinin,~S.~V.; Joy,~D.~C.;
  Xiao,~K.; Belianinov,~A.; Ovchinnikova,~O.~S. Nanoforging Single Layer MoSe2
  Through Defect Engineering with Focused Helium Ion Beams. \emph{Scientific
  reports} \textbf{2016}, \emph{6}, 30481\relax
\mciteBstWouldAddEndPuncttrue
\mciteSetBstMidEndSepPunct{\mcitedefaultmidpunct}
{\mcitedefaultendpunct}{\mcitedefaultseppunct}\relax
\EndOfBibitem
\bibitem[{El Abbassi} \latin{et~al.}(2021){El Abbassi}, Overbeck, Braun,
  Calame, {van der Zant}, and Perrin]{ElAbbassi.2021}
{El Abbassi},~M.; Overbeck,~J.; Braun,~O.; Calame,~M.; {van der Zant},~H.
  S.~J.; Perrin,~M.~L. Benchmark and application of unsupervised classification
  approaches for univariate data. \emph{Communications Physics} \textbf{2021},
  \emph{4}, 85\relax
\mciteBstWouldAddEndPuncttrue
\mciteSetBstMidEndSepPunct{\mcitedefaultmidpunct}
{\mcitedefaultendpunct}{\mcitedefaultseppunct}\relax
\EndOfBibitem
\bibitem[Eckmann \latin{et~al.}(2012)Eckmann, Felten, Mishchenko, Britnell,
  Krupke, Novoselov, and Casiraghi]{Eckmann.2012}
Eckmann,~A.; Felten,~A.; Mishchenko,~A.; Britnell,~L.; Krupke,~R.;
  Novoselov,~K.~S.; Casiraghi,~C. Probing the nature of defects in graphene by
  Raman spectroscopy. \emph{Nano letters} \textbf{2012}, \emph{12},
  3925--3930\relax
\mciteBstWouldAddEndPuncttrue
\mciteSetBstMidEndSepPunct{\mcitedefaultmidpunct}
{\mcitedefaultendpunct}{\mcitedefaultseppunct}\relax
\EndOfBibitem
\bibitem[Bae \latin{et~al.}(2013)Bae, Li, Aksamija, Martin, Xiong, Ong,
  Knezevic, and Pop]{Bae.2013}
Bae,~M.-H.; Li,~Z.; Aksamija,~Z.; Martin,~P.~N.; Xiong,~F.; Ong,~Z.-Y.;
  Knezevic,~I.; Pop,~E. Ballistic to diffusive crossover of heat flow in
  graphene ribbons. \emph{Nature communications} \textbf{2013}, \emph{4},
  1734\relax
\mciteBstWouldAddEndPuncttrue
\mciteSetBstMidEndSepPunct{\mcitedefaultmidpunct}
{\mcitedefaultendpunct}{\mcitedefaultseppunct}\relax
\EndOfBibitem
\bibitem[{El Sachat} \latin{et~al.}(2019){El Sachat}, K{\"o}enemann, Menges,
  {Del Corro}, Garrido, {Sotomayor Torres}, Alzina, and
  Gotsmann]{ElSachat.2019}
{El Sachat},~A.; K{\"o}enemann,~F.; Menges,~F.; {Del Corro},~E.;
  Garrido,~J.~A.; {Sotomayor Torres},~C.~M.; Alzina,~F.; Gotsmann,~B. Crossover
  from ballistic to diffusive thermal transport in suspended graphene
  membranes. \emph{2D Materials} \textbf{2019}, \emph{6}, 025034\relax
\mciteBstWouldAddEndPuncttrue
\mciteSetBstMidEndSepPunct{\mcitedefaultmidpunct}
{\mcitedefaultendpunct}{\mcitedefaultseppunct}\relax
\EndOfBibitem
\bibitem[Vakulov \latin{et~al.}(2020)Vakulov, Gireesan, Swinkels, Chavez,
  Vogelaar, Torres, Campo, de~Luca, Verheijen, Koelling, Gagliano, Haverkort,
  Alvarez, Bobbert, Zardo, and Bakkers]{Vakulov.2020}
Vakulov,~D. \latin{et~al.}  Ballistic Phonons in Ultrathin Nanowires.
  \emph{Nano letters} \textbf{2020}, \emph{20}, 2703--2709\relax
\mciteBstWouldAddEndPuncttrue
\mciteSetBstMidEndSepPunct{\mcitedefaultmidpunct}
{\mcitedefaultendpunct}{\mcitedefaultseppunct}\relax
\EndOfBibitem
\bibitem[Fugallo \latin{et~al.}(2014)Fugallo, Cepellotti, Paulatto, Lazzeri,
  Marzari, and Mauri]{Fugallo.2014}
Fugallo,~G.; Cepellotti,~A.; Paulatto,~L.; Lazzeri,~M.; Marzari,~N.; Mauri,~F.
  Thermal conductivity of graphene and graphite: collective excitations and
  mean free paths. \emph{Nano letters} \textbf{2014}, \emph{14},
  6109--6114\relax
\mciteBstWouldAddEndPuncttrue
\mciteSetBstMidEndSepPunct{\mcitedefaultmidpunct}
{\mcitedefaultendpunct}{\mcitedefaultseppunct}\relax
\EndOfBibitem
\bibitem[Cepellotti \latin{et~al.}(2015)Cepellotti, Fugallo, Paulatto, Lazzeri,
  Mauri, and Marzari]{Cepellotti.2015}
Cepellotti,~A.; Fugallo,~G.; Paulatto,~L.; Lazzeri,~M.; Mauri,~F.; Marzari,~N.
  Phonon hydrodynamics in two-dimensional materials. \emph{Nature
  communications} \textbf{2015}, \emph{6}, 6400\relax
\mciteBstWouldAddEndPuncttrue
\mciteSetBstMidEndSepPunct{\mcitedefaultmidpunct}
{\mcitedefaultendpunct}{\mcitedefaultseppunct}\relax
\EndOfBibitem
\bibitem[Simoncelli \latin{et~al.}(2020)Simoncelli, Marzari, and
  Cepellotti]{Simoncelli.2020}
Simoncelli,~M.; Marzari,~N.; Cepellotti,~A. Generalization of Fourier's Law
  into Viscous Heat Equations. \emph{Physical Review X} \textbf{2020},
  \emph{10}, 66\relax
\mciteBstWouldAddEndPuncttrue
\mciteSetBstMidEndSepPunct{\mcitedefaultmidpunct}
{\mcitedefaultendpunct}{\mcitedefaultseppunct}\relax
\EndOfBibitem
\bibitem[Luo \latin{et~al.}(2015)Luo, Maassen, Deng, Du, Garrelts, Lundstrom,
  Ye, and Xu]{Luo.2015}
Luo,~Z.; Maassen,~J.; Deng,~Y.; Du,~Y.; Garrelts,~R.~P.; Lundstrom,~M.~S.;
  Ye,~P.~D.; Xu,~X. Anisotropic in-plane thermal conductivity observed in
  few-layer black phosphorus. \emph{Nature communications} \textbf{2015},
  \emph{6}, 8572\relax
\mciteBstWouldAddEndPuncttrue
\mciteSetBstMidEndSepPunct{\mcitedefaultmidpunct}
{\mcitedefaultendpunct}{\mcitedefaultseppunct}\relax
\EndOfBibitem
\bibitem[Kang \latin{et~al.}(2017)Kang, Wu, and Hu]{Kang.2017}
Kang,~J.~S.; Wu,~H.; Hu,~Y. Thermal Properties and Phonon Spectral
  Characterization of Synthetic Boron Phosphide for High Thermal Conductivity
  Applications. \emph{Nano letters} \textbf{2017}, \emph{17}, 7507--7514\relax
\mciteBstWouldAddEndPuncttrue
\mciteSetBstMidEndSepPunct{\mcitedefaultmidpunct}
{\mcitedefaultendpunct}{\mcitedefaultseppunct}\relax
\EndOfBibitem
\bibitem[Islam \latin{et~al.}(2018)Islam, {van den Akker}, and
  Feng]{Islam.2018}
Islam,~A.; {van den Akker},~A.; Feng,~P. X.-L. Anisotropic Thermal Conductivity
  of Suspended Black Phosphorus Probed by Opto-Thermomechanical Resonance
  Spectromicroscopy. \emph{Nano letters} \textbf{2018}, \emph{18},
  7683--7691\relax
\mciteBstWouldAddEndPuncttrue
\mciteSetBstMidEndSepPunct{\mcitedefaultmidpunct}
{\mcitedefaultendpunct}{\mcitedefaultseppunct}\relax
\EndOfBibitem
\end{mcitethebibliography}

\end{document}